%% file: main_IEEE.tex
\documentclass[conference]{IEEEtran}
\IEEEoverridecommandlockouts
\usepackage{cite}
\usepackage{amsmath,amssymb,amsfonts}
\usepackage{algorithmic}
\usepackage{graphicx}
\usepackage{textcomp}
\usepackage{xcolor}
\def\BibTeX{{\rm B\kern-.05em{\sc i\kern-.025em b}\kern-.08em
    T\kern-.1667em\lower.7ex\hbox{E}\kern-.125emX}}
\begin{document}

\title{Team-oriented Consistency Checking of Heterogeneous Engineering Artifacts\\
\thanks{The research reported in this paper has been partly funded by FWF grants (P31989-N31 \& I4744-N), the Austrian COMET K1-Centre Pro2Future and the Austrian COMET K2-Center LCM, and the LIT Secure and Correct Systems Lab sponsored by the province of Upper Austria.}
}

\author{\IEEEauthorblockN{1\textsuperscript{st} Michael Alexander Tr\"ols}
\IEEEauthorblockA{\textit{Johannes Kepler University}\\
Linz, Austria \\
michael.troels@jku.at}
\and
\IEEEauthorblockN{2\textsuperscript{nd} Atif Mashkoor}
\IEEEauthorblockA{\textit{Johannes Kepler University}\\
Linz, Austria \\
atif.mashkoor@jku.at}
\and
\IEEEauthorblockN{3\textsuperscript{rd} Alexander Egyed}
\IEEEauthorblockA{\textit{Johannes Kepler University}\\
Linz, Austria \\
alexander.egyed@jku.at}
}

\maketitle

\begin{abstract}
\input{sections/abstract}
\end{abstract}


\input{main_body}

\bibliographystyle{acm}
\bibliography{bibliography}

\end{document}

%% file: sections/abstract.tex
Consistency checking of interdependent heterogeneous engineering artifacts, such as requirements, specifications, and code, is a challenging task in large-scale engineering projects. The lack of team-oriented solutions allowing a multitude of project stakeholders to collaborate in a consistent manner is thus becoming a critical problem. In this context, this work proposes an approach for team-oriented consistency checking of collaboratively developed heterogeneous engineering artifacts.

%% file: main_body.tex
\input{sections/introduction}

\input{sections/architecture}

\input{sections/methodology}

\input{sections/conclusion}

%% file: sections/introduction.tex
\section{Introduction}
The collaborative development of modern engineering projects is becoming increasingly complex. A multitude of project stakeholders produces an enormous amount of work results (i.e., engineering artifacts) that require consideration in a combined context. For example, requirements, specifications, code and other artifacts are strongly interdependent, thus changing one artifact requires the propagation of the change into several others. As a result, maintaining the consistency of heterogeneous artifacts becomes a daunting task in large-scale projects. In this regard, collaborative engineering environments, e.g., DesignSpace~\cite{Demuth15}, can offer help. While the works discussed in literature (e.g., \cite{troels2019, troels2019ME, sabetzadeh2008}) can handle global consistency checking for heterogeneous engineering artifacts, their employed workflow only supports the counter-checking of one engineer's work with another's (i.e., one-to-one consistency checking). This is due to the internal organization of the collaborative engineering environment's artifact storage. This storage is structured in hierarchical work areas, which only allows the consistency checking of a child work area with a parent work area. A team-oriented engineering effort, however, demands a similar team-oriented consistency checking workflow (i.e., one-to-many consistency checking). Though this endeavor is an ongoing research challenge~\cite{troels2019EICS, troels2020JSME}. 

The work presented in this paper focuses on employing global consistency checking in a team-oriented context. Engineers synchronize their work with a collaborative engineering environment, where the contents of their individual work areas can automatically be counter-checked against the contents of a pre-defined group of other work areas. This way, engineers can check the consistency of their work in the context of their respective team members' work. They can benefit from immediate consistency checking feedback, without being dependent on a hierarchy of work areas that determines their perspective on the artifact storage.

%% file: sections/architecture.tex
\section{Architecture}
\label{sec:architecture}
In this work, we utilize the collaborative engineering environment DesignSpace~\cite{Demuth15} and extend it in terms of its team-oriented consistency checking capabilities. DesignSpace features a readily-available consistency checker along with the possibility to synchronize engineering artifacts through network. To prepare engineering artifacts for this environment, a custom adapter transforms an engineering tools' internal data objects into a uniform data representation. In this representation, artifacts are considered as a mapping between different properties and their respective changes (e.g., a requirement's artifact may contain a property for the requirement's name -- the latest change value on this property represents the current name of the requirement). The tool adapter then synchronizes these changes with an artifact storage. A consistency checker can then feasibly analyse artifacts according to a pre-defined set of global consistency rules in a single centralized location. In the artifact storage, changes are persisted within individual work areas. Every tool adapter is linked to its own work area. The work areas are organized in a hierarchy through which changes are committed upwards step-by-step. All sub-ordinate work areas can retrieve the changes from a super-ordinate work area. Retrieving an artifact from the lowest work area in the hierarchy retrieves the said artifact with all changes that are stored within that specific work area along with all changes from super-ordinate work areas. In case of overlaps (i.e., changes on the same properties), the changes from the lower work areas overrule the changes from the higher ones. This way, engineers retrieving artifacts from their work area gain a perspective on the artifact storage individualized by their work areas' specific changes as well as the changes stored within the super-ordinate work areas. 

For different sub trees of the hierarchy, the same artifacts may be represented in an entirely different way, because the corresponding work areas contain different changes. This concept is comparable to Git branches~\cite{loeliger2012version}. 
However, it is based on the changes of an artifact's properties directly integrated with the ongoing changes of higher work areas, rather than an altered base version of an artifact that is merged with its original later. 
The property-based nature of the utilized collaborative engineering environment further allows for partial synchronization of artifacts (e.g., only the method headers of a code class). In comparison to current collaboration systems like Git -- where the entirety of a file has to be synchronized and merged -- this is relatively lightweight.
This lightweight synchronization practically allows for continuous live communication and feedback between the custom tool adapters and the environment. This capability is further supported by the integration of tool adapters into the tools' APIs, allowing a direct synchronization of a tool's internal data objects. This stands in contrast with the usual file-based synchronization of artifacts, as it is the case in Git and other modern collaboration systems.

%% file: sections/methodology.tex
\section{Approach}
\label{sec:approach}
In the proposed approach, the engineers first pick a group of work areas with which they wish to counter-check their changes.
Usually the collaborative engineering environment's consistency checker retrieves an artifact from a specific work area, thus gaining the same perspective as the engineer's, and then counter-checks it with whatever changes are available from that perspective (i.e., a consistency check is always hierarchy-dependent). The results are then written back into the work area in the form of a dedicated engineering artifact. 

\begin{figure}[b]
  \centering
    \includegraphics[width=0.5\textwidth]{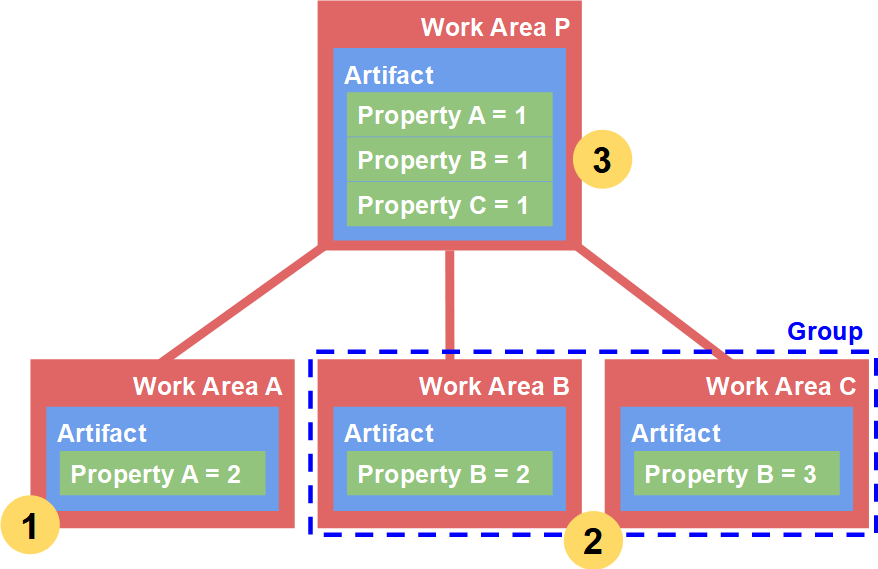}
    \caption{Hierarchical work areas and the numbered search order for properties in team-oriented consistency checking}
    \label{fig:searchorder}
\end{figure}

In team-oriented consistency checking, the retrieval of artifacts must be re-routed through the pre-defined group of work areas, before it falls back on the hierarchy. In other words, the consistency checker first retrieves changes from the engineer's work area. Next it retrieves changes from the pre-defined group of work areas and eventually from the hierarchy super-ordinate to the engineer's work area. This search order is illustrated in Figure \ref{fig:searchorder}. To check the consistency on the artifact, the consistency checker must first retrieve the property changes from work area A, then the correct changes from the group, and then the last property -- property C, which is neither available in work area A, nor in one of the grouped work areas -- from work area P. Within the group there may be overlapping changes on the same property. In that case, the retrieval mechanism decides on the basis of a timestamp that is assigned to each change at the point of its synchronization.
With the sum of all retrieved properties a full artifact can be constructed for consistency checking. The artifacts respective values are substituted in the currently evaluated consistency rule. Results are written back into the work area that triggered the consistency check (in the illustrative case, work area A).

Team-oriented consistency checks are not persisted in the same way as hierarchy-specific consistency checks. In the latter case, the result must match the consistency state of the direct super-ordinate work area, under the assumption that the changes from the checked work area were committed there (i.e., consistency results that are stored within work area A must also be true for work area P once A commits its work to P). Instead, results are recorded in a work area specific group result property. By doing so, the artifact dedicated to storing consistency results can record all team-oriented consistency checks and retain the regular hierarchy-specific results.

%% file: sections/conclusion.tex
\section{Conclusion}
In this paper, we introduced an approach of consistency checking among heterogeneous engineering artifacts, under the consideration of pre-defined teams. The proposed approach utilizes the concept of global consistency checking and an existing collaborative engineering environment. As a result, we can check the consistency of an engineers' work within a team environment, without merging the engineering artifacts or interrupting the workflow.